\documentclass[twoside]{dis04}

\usepackage{epsfig}  

\def\runauthor{Arnold Pompo\v s}
\def\shorttitle{Other Particle Searches at the Tevatron}

\def \invpb    {\relax\ifmmode{\rm pb^{-1}}\else{$\rm pb^{-1}$}\fi}

\def\Journal#1#2#3#4{{#1} {\bf #2}, #3 (#4)}

\def\NIM{\em Nucl. Instrum. Methods}

\def\NPB{{\em Nucl. Phys.} B}
\def\PLB{{\em Phys. Lett.}  B}
\def\PRL{\em Phys. Rev. Lett.}
\def\PRD{{\em Phys. Rev.} D}

\def\dz{D\O\,\,}
\def\pt{$p_{\rm T}\,\,$}
\def\et{$E_{\rm T}\,\,$}
\def\gvcc{GeV/$c^{2}$\,\,}
\def\gevc{GeV/$c$\,\, }
\def\tvcc{TeV/$c^{2}$\,\,}

\newcommand{\met}{\mbox{${\hbox{$E$\kern-0.6em\lower-.1ex\hbox{/}}}_{\rm T}$}}

\begin{document}

\title{Other Particle Searches at the Tevatron
}

\author{Arnold Pompo\v s}

\address{
(On behalf of the CDF and the \dz collaborations)\\
\vspace*{0.3cm}
The University of Oklahoma \\
\dz Experiment, MS 352, Fermilab\\
Batavia, IL 60510, USA\\
E-mail: pompos@fnal.gov }

\maketitle

\abstracts{The CDF and \dz collider experiments at Fermilab have searched 
for evidence of physics extending 
beyond the scope of the Standard Model of particles and interactions. We report
on the results of searches for heavy extra gauge bosons, extra dimensions and composite electrons
in 200 \invpb (per experiment) of $p\bar p$ collisions 
at $\sqrt{s}$ = 1.96 TeV. No evidence 
of a new signal has been found, therefore limits on model parameters have been derived.}

\section{Introduction}

The Standard Model of particles and interactions (SM) 
is by far the most successful theory 
describing high-energy physics data. It's predictions have been verified to great accuracy, but 
many fundamental questions still remain unanswered. 
Among them are the mechanism of electroweak symmetry breaking, why the gravitational interaction seems 
to be so weak in comparison to the other fundamental interactions and whether the fundamental interactions 
unify into one underlying physical interaction. Several models such as Large Extra Dimensions (LED), 
Grand Unified Theories (GUT), Supersymmetry (SUSY) and Technicolor (TC) have been proposed to address such questions.
These models extend beyond the scope of the Standard Model and predict new signatures that can be detected at current high-energy colliders such as 
the Tevatron\footnote{$p\bar p$ collider at $\sqrt{s}$ = 1.96 TeV located in the vicinity of Chicago in the USA. It expects to deliver $\approx$ 8000 \invpb~of Run~II data until the end of 2009. The Tevatron hosts two collider experiments: The Collider Detector at Fermilab (CDF)~\cite{CDF} and the \dz Experiment~\cite{D0}. }. Searches for new phenomena though face some experimental challenges. The production cross sections tend to be very small (in the range of 1 pb) whereas the deep inelastic $p\bar p$ cross section is of the order of $10^{10}$ pb. It is often hard to distinguish new phenomena signal from known SM background processes. Jet based strategies are overwhelmed by SM processes, so we employ lepton based signatures, even though the rates are often suppressed.
 The results shown at this conference 
are based on 200 \invpb~per experiment of Run~II delivered in 2002 and 2003.

\section{Search for Extra Dimensions}

Various models of string theory suggest the existence of additional, finite size (R),  
spatial dimensions to the ones we sense in our everyday lives. If extra dimensions of large size exist, they could play an important role explaining 
why in our 3+1 dimensions the gravitational interaction is so much weaker than the 
other fundamental interactions.

\begin{figure}[t!]
\begin{center}
\begin{minipage}{0.4\linewidth}
\epsfig{figure=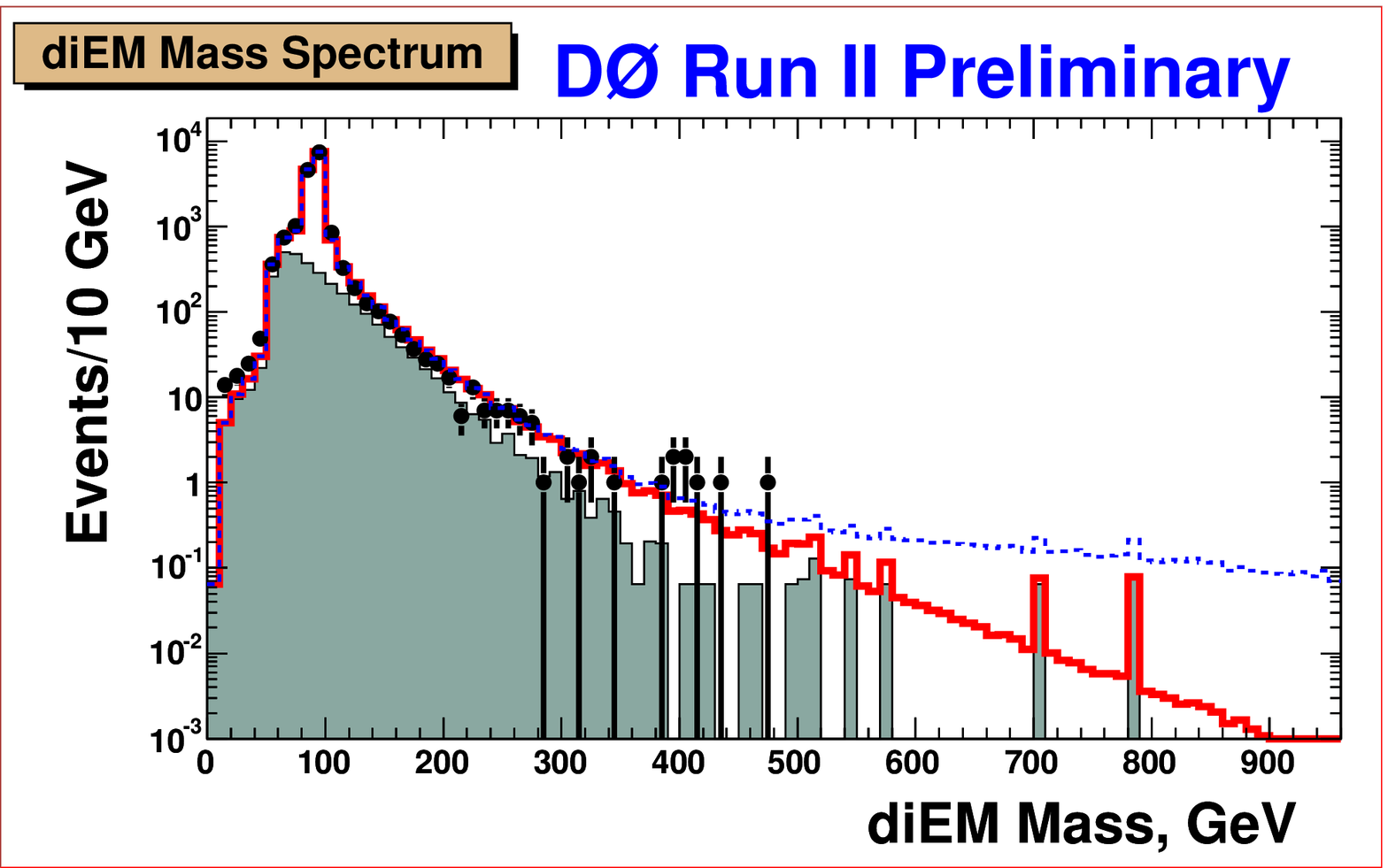,height=1.8 in}
\end{minipage}
\hspace*{2.3cm}
\begin{minipage}{0.4\linewidth}
\epsfig{figure=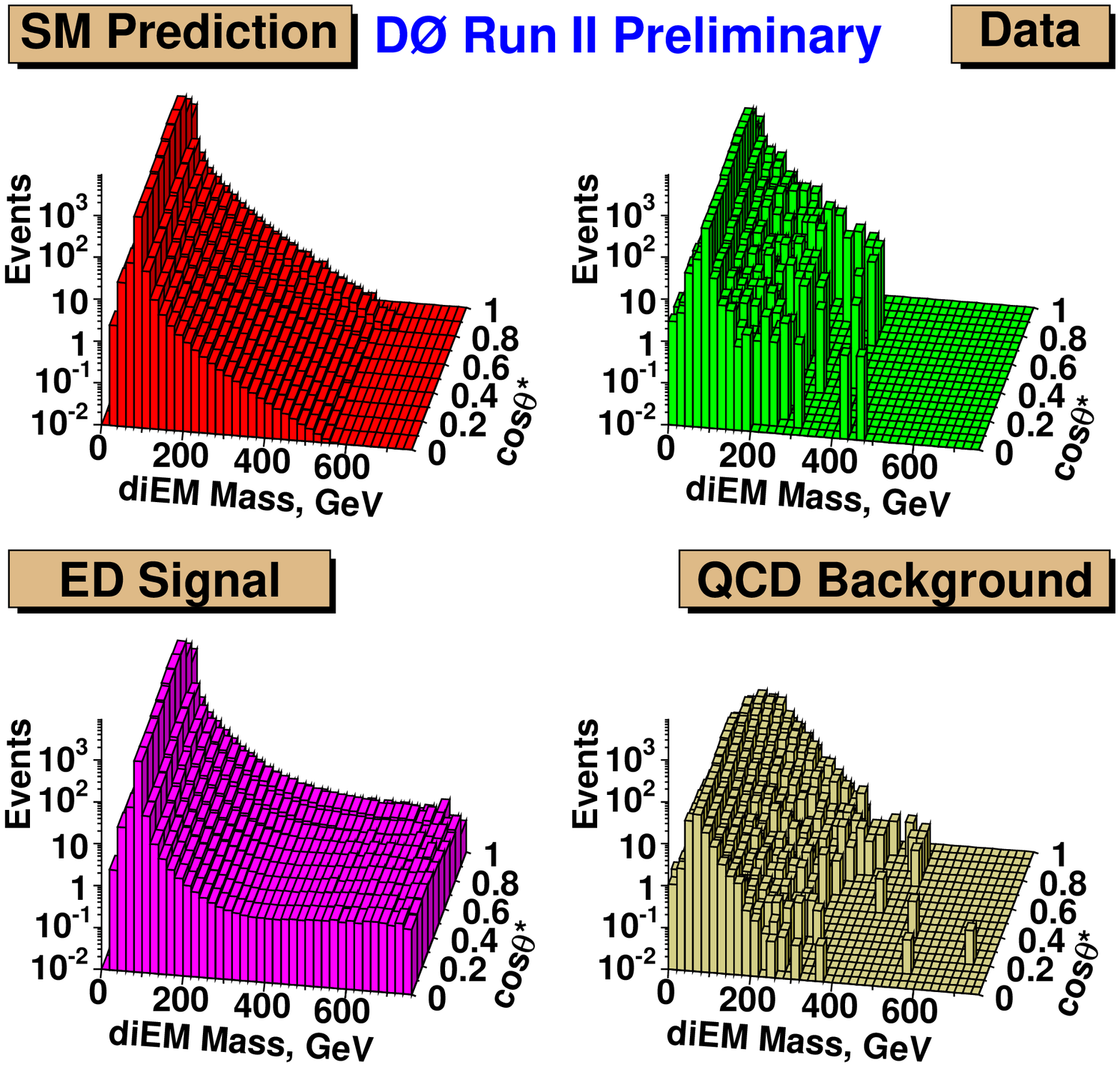,height=1.8 in}
\end{minipage}
\caption[*]{Left: The di-em mass distribution at \dz. Points are data, light filled histogram is QCD background, solid
line is a fit to total background. Dashed line is an example of ED signal. Right: di-em mass vs. cos(scattering angle) 
distributions for SM background (top left), data (top right), ED signal plus background (bottom left) and QCD background (bottom right).}
\label{fig:led}
\end{center}
\end{figure}

\subsection{ADD Model for Large Extra Dimensions}

In the ADD~\cite{ADD} model, the Standard Model's chiral fermions and gauge bosons 
are confined to a 
3-dimensional membrane. At the same time, gravity is allowed to propagate 
in all 3+n spatial dimensions. Fundamentally, gravity is as strong 
as the other gauge interactions ($M_{PL}(3+{\rm n})\approx 1$ \tvcc), but this is apparent only for a 3+n dimensional observer. For a 3 dimensional observer, gravity is volume suppressed ($M_{PL}(3{\rm D})\approx 10^{16}$ \tvcc) and its strength depends on the n-th power of the 
compactification
 radius R. While tabletop gravity experiments and astrophysical observations started to produce tight limits on the size of R and the number n, for ${\rm n}\ge 3$ colliders are the only sensitive probe. There are two main ways of probing large extra dimensions at colliders: 1) To look at the effects of virtual graviton exchange in fermion or boson pair production, 2) To look for the production of a real 
graviton       
recoiling against a gauge boson or a quark.  In both of these cases, the strength of the effect depends on $M_{PL}(3+{\rm n})$ allowing to test R and n or set limits on $M_{PL}(3+{\rm n})$.

The first type of search for LED mentioned above focuses on the very high mass Drell-Yan region (due to heavy KK modes of the graviton).
The expected signal has very energetic electrons and photons, therefore the very efficient ($>$ 99\%) 
 high-\pt single electron, dielectron or diphoton
triggers were applied at \dz. Events with central-central or central-forward, isolated, \et $>$ 25 GeV e's and $\gamma$'s were selected.
No tracking requirement has been made to treat the dielectron sample the same way as the diphoton one. Drell-Yan and dijet QCD (with EM-like
jets) are the main background sources. They were normalized by fitting to the low di-em mass spectrum where no LED signal is expected. Figure~\ref{fig:led} shows the di-em invariant mass spectrum of \dz. CDF explored its dielectron sample, with similar selection as \dz, except a track requirement for electrons. Figure~\ref{fig:cdfdilepton} shows CDF's dielectron invariant mass distribution. 

Since no signal excess above expected background has been observed, \dz extracts limits by using a 2-D fit to the dilepton invariant mass and scattering angle distributions while CDF fits the dielectron invariant mass distribution. Both CDF and \dz set limits on $M_{PL}(3+{\rm n})$. Three formalisms were considered, GRW~\cite{grw}, HLZ~\cite{hlz} and Hewett~\cite{hew}.  
Within  the GRW formalism, CDF sets an upper limit $M_{PL}(3+{\rm n}) > 1.11$ \tvcc. \dz produced a slightly higher limit\footnote{In case of \dz, the MC simulation has been scaled up by 30\%
to estimate NLO cross sections.} $M_{PL}(3+{\rm n}) > 1.36$ \tvcc and when combined with data from 1992-1995 (called Run I), \dz finds  $M_{PL}(3+{\rm n}) > 1.43$ \tvcc.
In the HLZ formalism, CDF's upper limit on $M_{PL}(3+{\rm n})$ were: 1.11, 1.17, 0.99, 0.89, 0.83 and 0.79~\tvcc for n= 2 ... 7 respectively. \dz's limits based on combined data with Run I in the HLZ formalism were: 1.67, 1.7, 1.43, 1.29, 1.20 and 1.14 \tvcc for n as above.
In the Hewett formalism, CDF sets 0.99 and 0.96 \tvcc limit for $\lambda = +1$ and -1 respectively, while \dz sets 1.22 and 1.10 \tvcc. 

\dz also used 100 \invpb~of isolated, \pt $>$ 15 \gevc dimuon data to search for LED. They set $M_{PL}(3+{\rm n}) > 0.88$ \tvcc limit in he GRW formalism, 0.75, 1.05, 0.88, 0.80, 0.74 and 0.70 \tvcc in the HLZ formalism for n = 2 ... 7 and 0.79~\tvcc in the Hewett formalism for positive $\lambda$.

The second type search for LED, where a real graviton is produced in the collisions, escapes into the bulk, and leaves behind a monojet 
 was carried out at \dz using  85 \invpb~of jet data. 
Events with one central jet of \et $>$ 150 GeV and \met $>$ 150 GeV separated by at least 
 $30^{\circ}$ in $\phi$ were selected. 
(Leptons were vetoed and the second jet's \et was required to be $<$ 50 GeV in order to reduce background.)
The main (and irreducible background) comes from $Z\to \nu\nu$ + 1 or 2 jets. They observe 63 events while expecting 100 but with 60\% 
uncertainty which mostly comes from large jet energy scale uncertainties. \dz sets an upper limit of  $M_{PL}(3+{\rm n})$ being $\approx$ 700 \gvcc across the range of extra dimensions of 4 to 7. 

\subsection{Randall-Sundrum Model}

In the Randal-Sundrum (RS) Model~\cite{RS}, one extra dimension which is compactified to radius R is assumed. The weakness of the 3D gravity is explained by an exponential suppression which depends on the effective Planck scale 
$M_{PL}(eff)$, radius R and on a scale factor $k$ of the order of the $M_{PL}(3{\rm D})$. At colliders, the RS scenario can be probed by looking at the effect graviton exchange has on  dilepton production, which depends on the ratio 
$k/M_{PL}(eff)$.  

\begin{figure}[t!]
\begin{center}
\begin{minipage}{0.4\linewidth}
\epsfig{figure=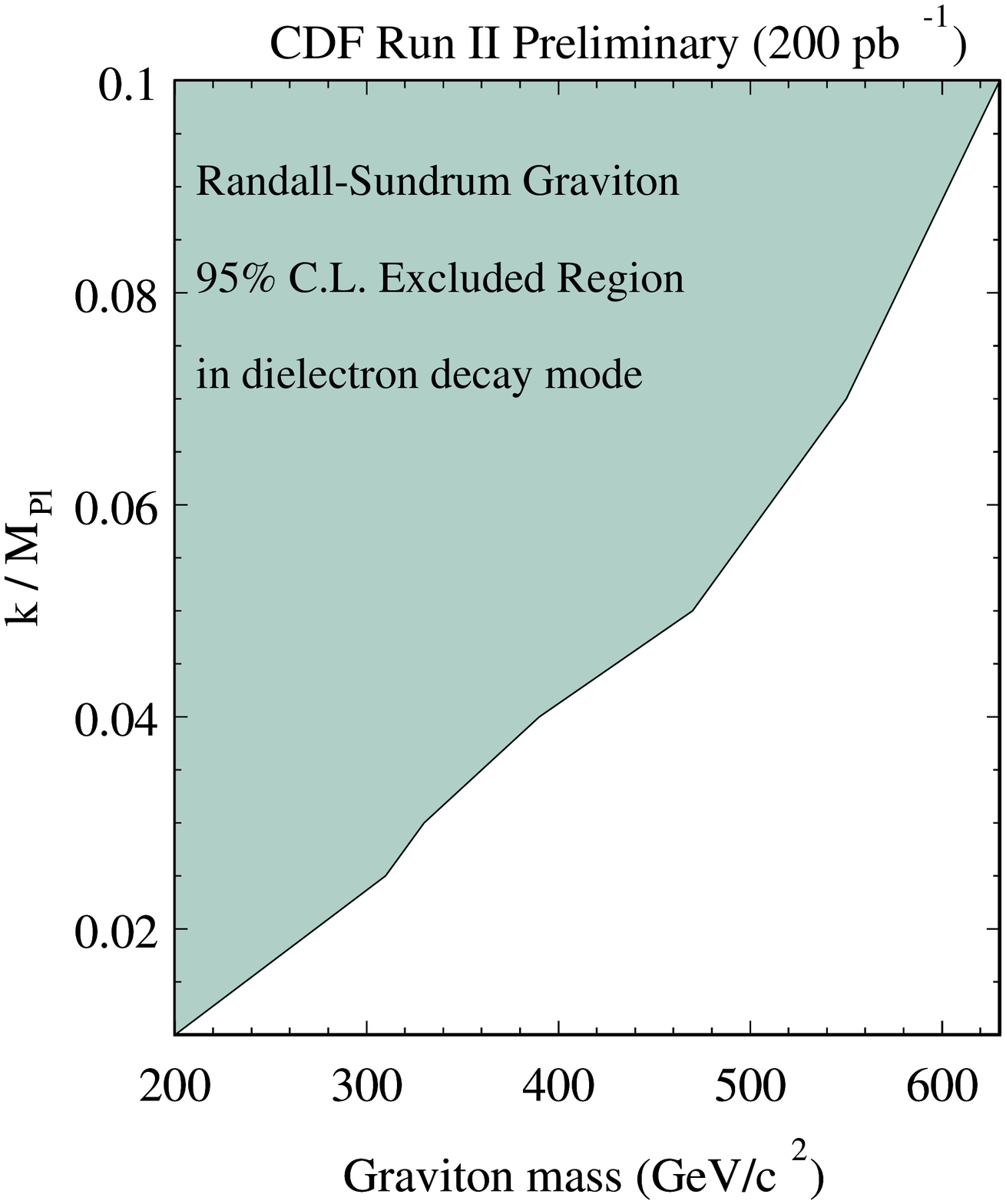,height=2.in}
\end{minipage}
\hspace*{0.5cm}
\begin{minipage}{0.4\linewidth}
\epsfig{figure=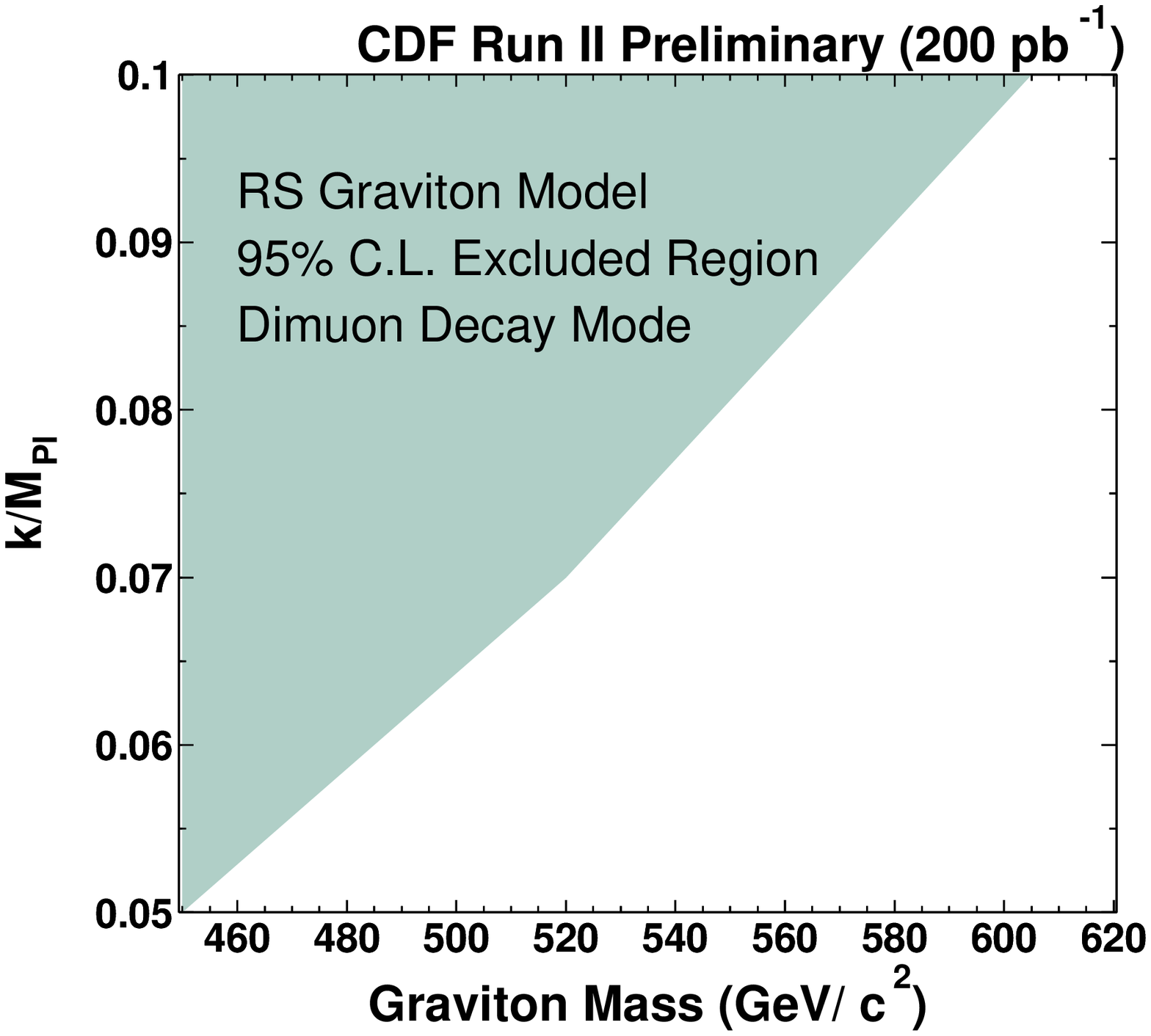,height=2.in}
\end{minipage}
\caption[*]{CDF's 95\% C.L. excluded region in $ee$ and $\mu\mu$ Randall-Sundrum model,}
\label{fig:rs}
\end{center}
\end{figure}

CDF has searched  its high-\pt dimuon and dielectron samples to search for RS extra dimension. The dielectron sample is the same as in case of LED described in the previous paragraph. The dimuon sample consists of events with two isolated, \pt $>$ 20 \gevc muons. Comparison of data with expected background is shown in Figure~\ref{fig:cdfdilepton}. Since no excess of data above expected SM background has been observed, CDF employs a likelihood fit to the invariant dilepton mass distributions and excludes regions up to $\approx$ 600 \gvcc in graviton mass for $k/M_{PL}(eff)= 0.1$ as shown in Figure~\ref{fig:rs}.

\subsection{${\rm TeV^{-1}}$ Extra Dimensions}

In this model~\cite{TEV} matter resides on a high dimensional p-brane with 
chiral fermions confined to the ordinary 3 spatial dimensions but allowing gauge bosons to propagate in the extra (compactified)  dimensions. This gives rise to Kaluza-Klein (KK) towers of the gauge bosons whose direct production or virtual effects would alter the dilepton production at high-energy colliders. (Gravity in the bulk is not of direct concern in this model.) The mass and the effect of the KK states depend on the number of extra spatial dimensions and on the compactification radius R~$\approx 1/M_C$. ($M_C$ is the compactification scale.) \dz performed a dedicated search in its 200 \invpb~of high-\pt 
dielectron sample (same sample as used in the ADD LED di-em search but electrons were not isolated and were 
required to have associated tracks.) By similar technique as in ADD LED search, \dz sets an upper limit of $M_C >1.12$ \tvcc.

\section{Heavy Gauge Boson Search}

Extra U(1) gauge bosons, $Z'$, are predicted by many extensions of the SM, such as the SO(10)~\cite{ZP1} or $E_6$ ( $Z_{I}$, $Z_{\psi},\,Z_{\chi},\,X_{\eta}$)~\cite{ZP2} based Grand Unification Theory. 
As a reference ``model'' for experimental comparisons, it is often useful to consider a $Z'$ boson whose coupling to fermions is Standard Model like. 
Experimentally, it is possible then to search for excess production of dilepton pairs at large invariant masses caused by exchange of real or virtual $Z'$. The main background events are coming from Drell-Yan events and QCD processes, such as misidentified jets. 

\begin{figure}[t!]
\begin{center}
\begin{minipage}{0.4\linewidth}
\epsfig{figure=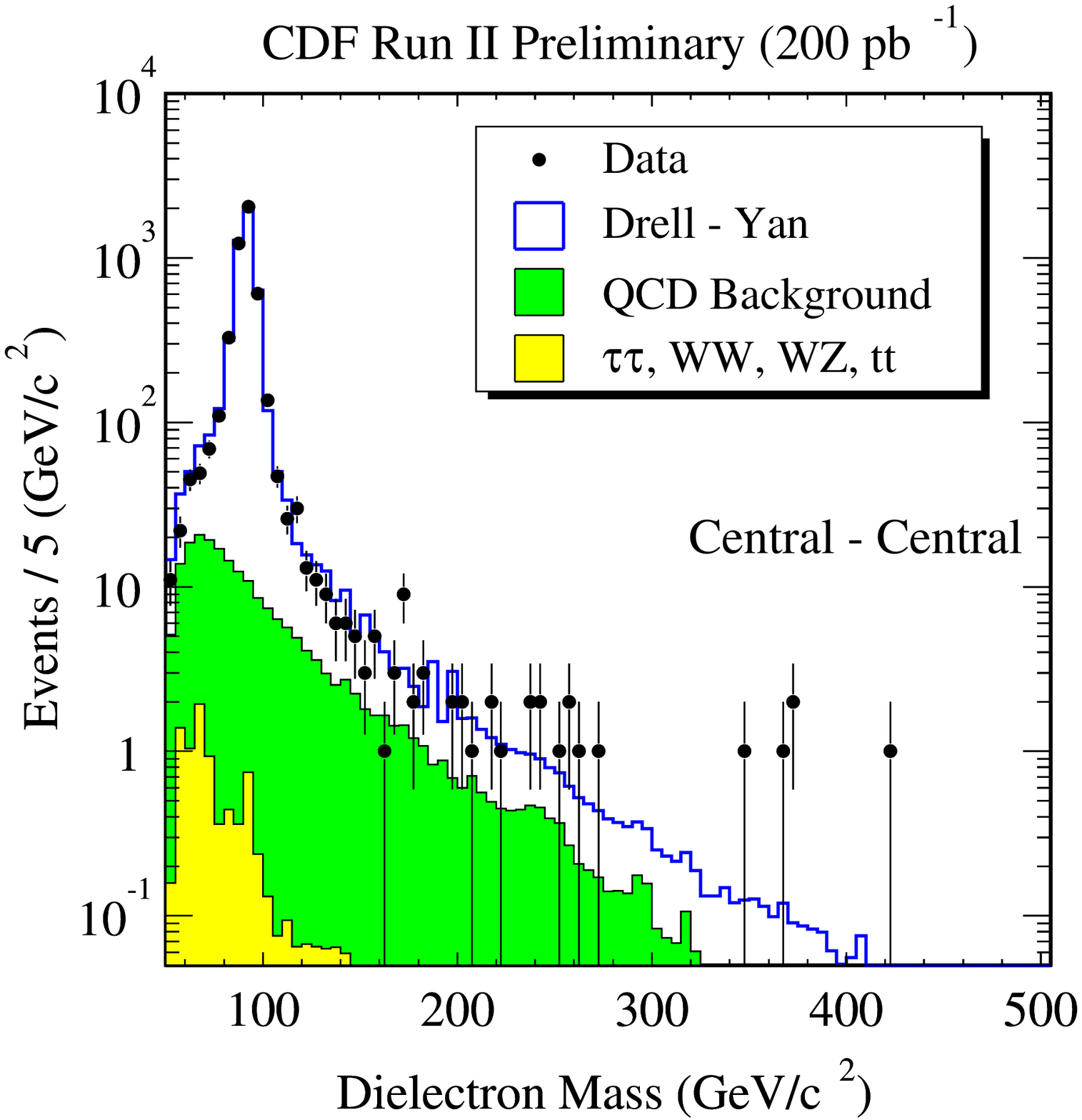,height=2.in}
\end{minipage}
\hspace*{0.5cm}
\begin{minipage}{0.4\linewidth}
\epsfig{figure=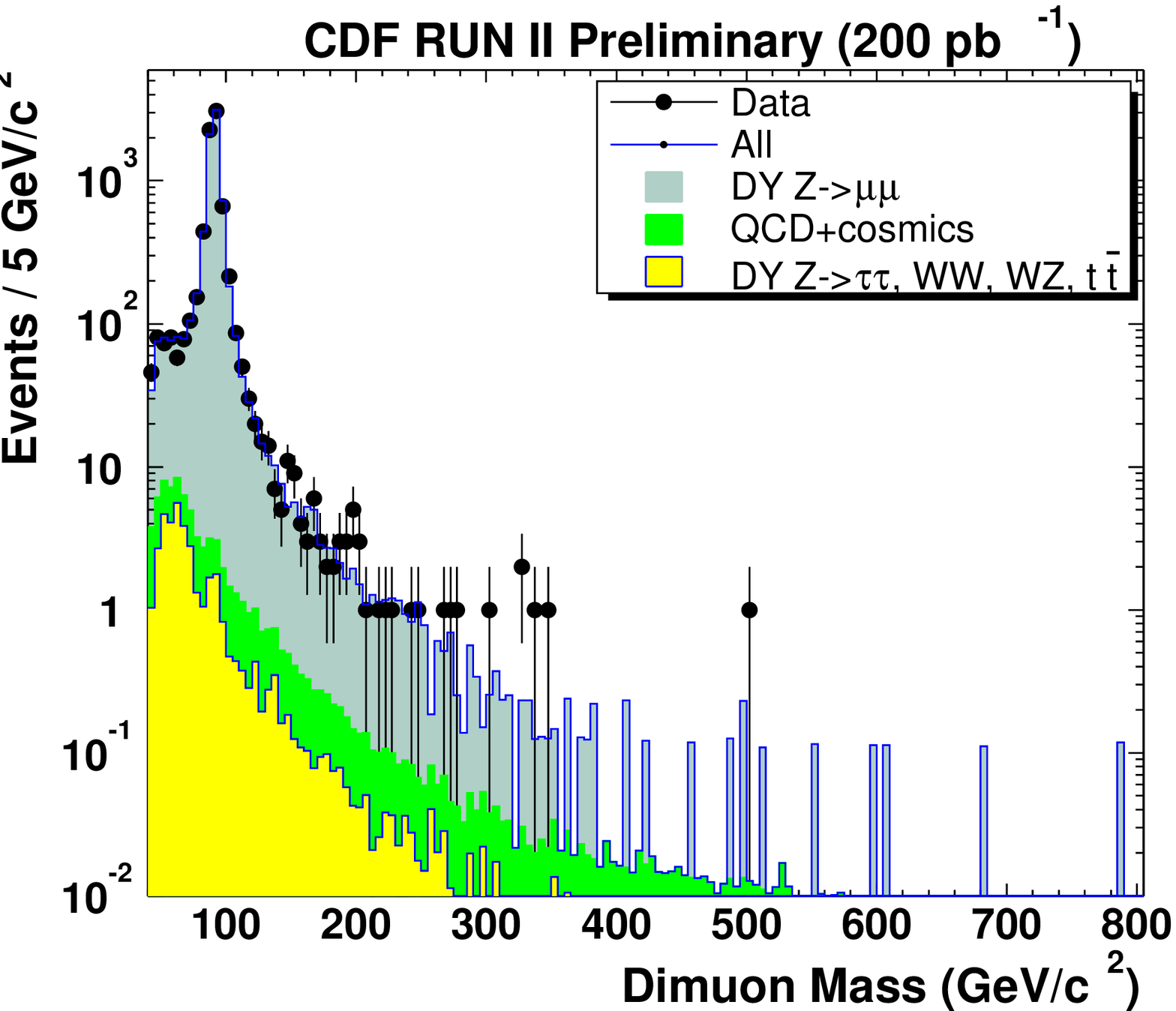,height=2.in}
\end{minipage}
\caption[*]{CDF's dielectron and dimuon invariant mass distributions. }
\label{fig:cdfdilepton}
\end{center}
\end{figure}

CDF utilized the same high-\pt dielectron and dimuon data sets as used in the RS LED search (Figure~\ref{fig:cdfdilepton}). 
\dz obtained $Z'$ search results on its high-\pt dielectron sample as used in the ${\rm TeV^{-1}}$ ED search.
No deviations from the SM predictions have been observed by neither experiments, therefore 95\% C.L. upper limits on $Z'$ masses were set. In the $ee$ channel CDF sets a limit of 
750, 570, 610, 625 and 650 \gvcc for $Z'$ with SM-like couplings, $Z_{I}$, $Z_{\chi},\,Z_{\psi},\,X_{\eta}$ respectively. In the $ee$ channel \dz obtained similar results of 780, 575, 640, 650 and 680 GeV in the same order as above. In the $\mu\mu$ channel CDF obtained upper limits 735, 530, 600, 635 and  580 \gvcc in the same order as above.    

\section{Search for Excited Electrons}

\begin{figure}[t!]
\begin{center}
\begin{minipage}{0.4\linewidth}
\epsfig{figure=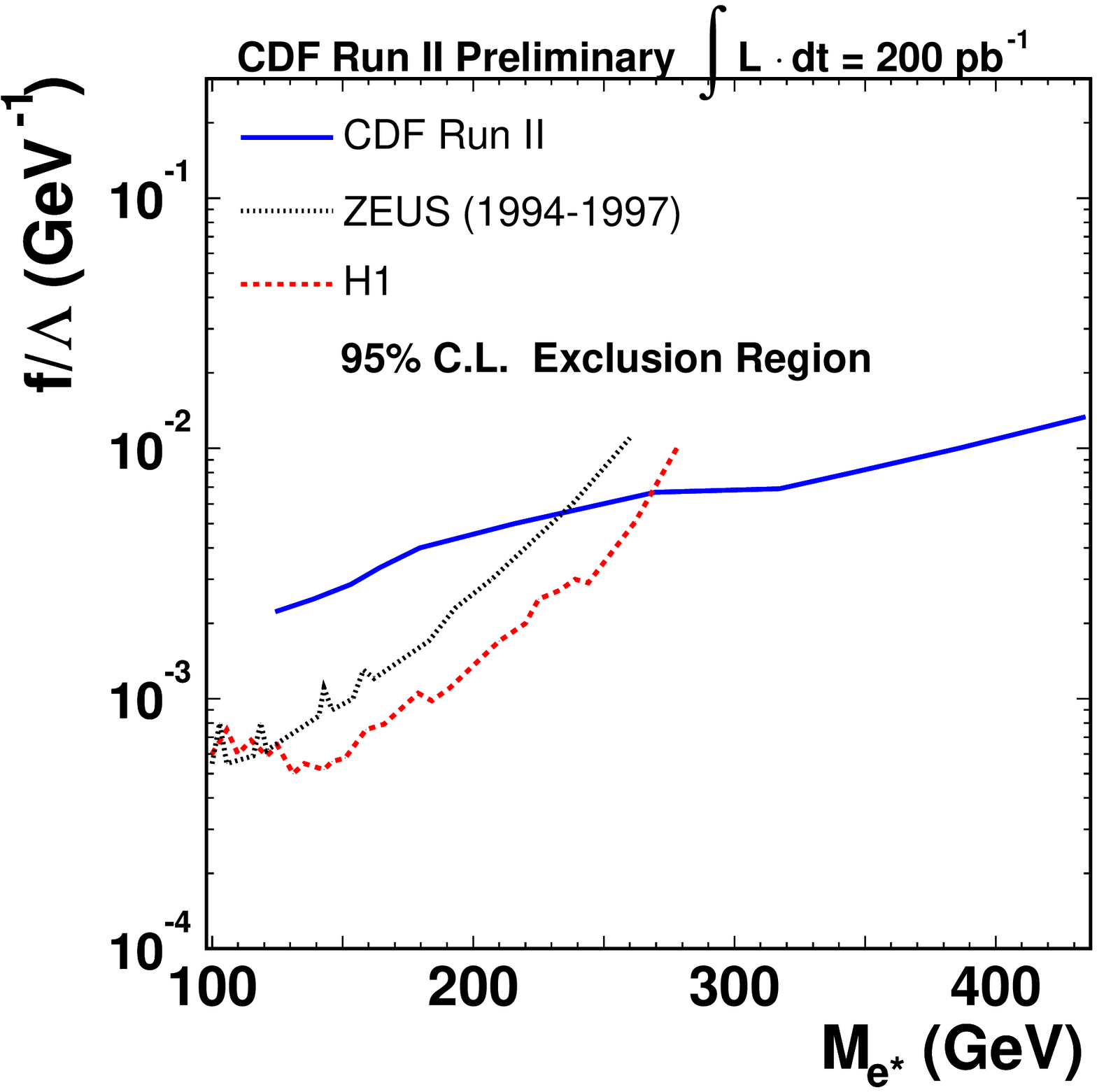,height=2.in}
\end{minipage}
\hspace*{0.5cm}
\begin{minipage}{0.4\linewidth}
\epsfig{figure=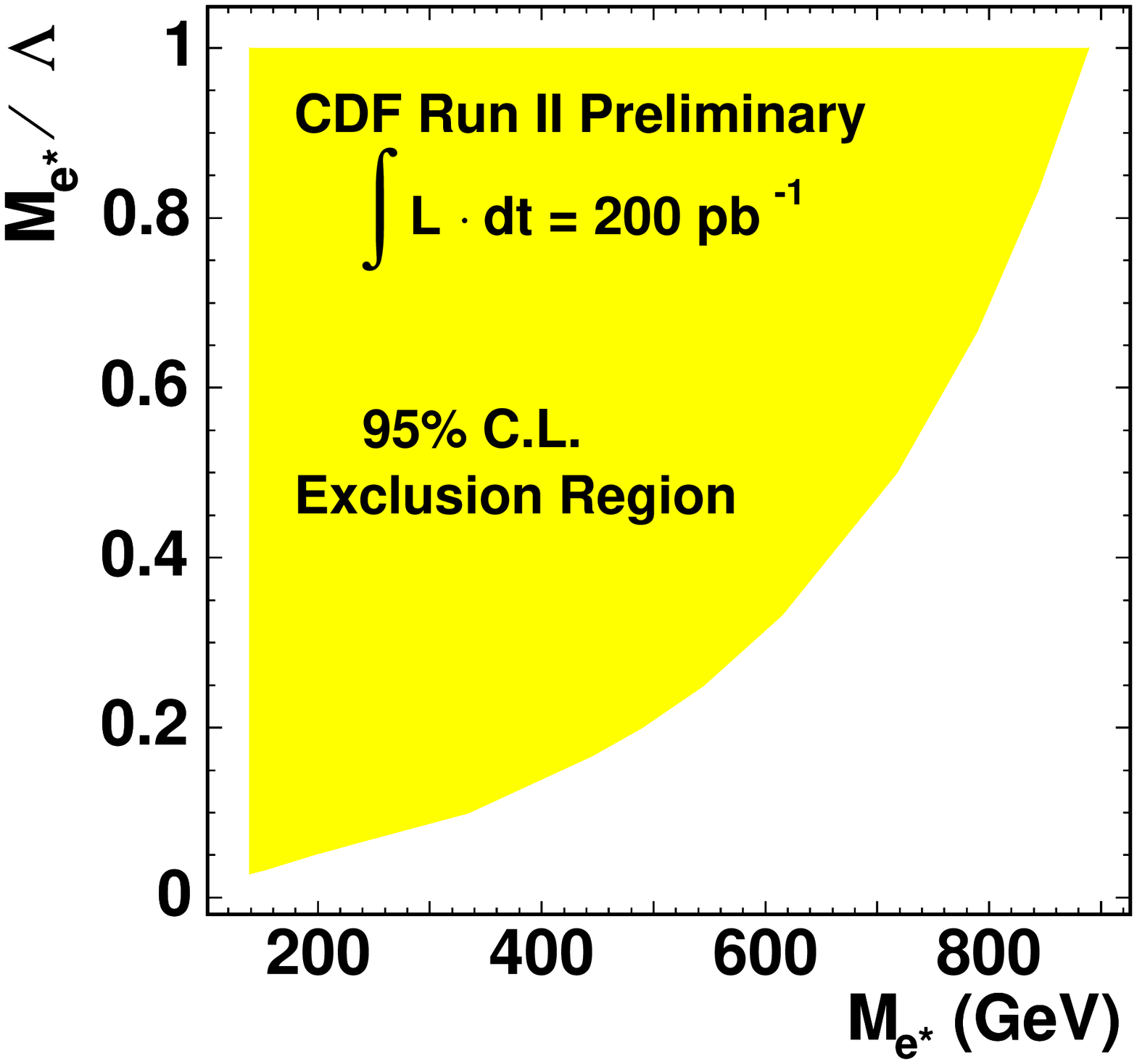,height=2.in}
\end{minipage}
\caption[*]{ CDF 95\% CL excluded region in the mass parameter versus
$f/\lambda$ for the gauge mediated model or versus $M_{e^*}/\Lambda$
for the contact interaction model.}
\label{fig:exe}
\end{center}
\end{figure}

One indication of a need for a theory going beyond the SM would be a discovery of
 excited leptons or quarks. At hadron colliders excited electrons, $e^*$, could be produced 
through contact interactions~\cite{EXCITEDE} or gauge mediated
interactions~\cite{EXCITEDE}. CDF has searched 200 \invpb~for excited electron-electron pair
production with $e^*$ 
decaying to an electron and  photon. The signature would therefore be events with ee$\gamma$ with an 
M(e$\gamma$) resonance. Events with two electrons and a photon with $E_T > 25$~GeV 
were selected. No access of events above the SM background of Z$\gamma$, Z+jets, WZ, QCD multijets and 
$\gamma\gamma$+jets has been observed. Upper limits are set at 95\%   
C.L. in the parameter space of M($e^*$)/$\Lambda$ vs M($e^*$)~\cite{EXCITEDE} in case of the contact interaction
model and of f/$\lambda$ vs M($e^*$)~\cite{EXCITEDE} in case of the gauge mediated
interaction model. Figure ~\ref{fig:exe} shows the
excluded region for the two models.  

\section{Summary} 

Tevatron's CDF and \dz collaborations analyzed their
 first 200 \invpb of high-\pt di-em and dimuon samples 
obtained from  $p\bar p$ collisions at $\sqrt{s}$ = 1.96 TeV.
No deviation of data from Standard Model predictions have been observed, 
and strong limits were set in terms of parameters of proposed extra dimensions, 
heavy gauge bosons and excited electron models.

\section*{Acknowledgments} 

I would like to thank the organizers of DIS2004, especially 
 Dr.\ Du\v san Bruncko, for creating this unforgettably warm atmosphere in such a beautiful place as the 
High Tatra mountains in Slovakia. Many thanks to the members of the CDF and \dz
New Phenomena working groups for achieving these great results presented at this conference. 
Stephan Lammel, thank you very much for your valuable comments and help with preparing my DIS2004
presentation.

\end{document}